\documentclass[12pt]{article}

\usepackage{latexsym}
\usepackage{epsfig}
\usepackage{amsfonts}
\newcommand{\sect}[1]{\setcounter{equation}{0}\section{#1}}

\def\be{\begin{equation}}
\def\ee{\end{equation}}
\def\ba{\begin{eqnarray}\samepage}
\def\ea{\end{eqnarray}}
\def\mr{\mathrm}
\def\mc{\mathcal}
\def\p{\phantom}
\def\definetobe{\stackrel{\mr{def}}{=}}

\begin{document}

\title{{\bf Dynamic and Thermodynamic Stability and Negative Modes in
Schwarzschild-Anti-de Sitter}}   
 
\author{\\Tim Prestidge \\ \\{\sl T.Prestidge@damtp.cam.ac.uk} \\ \\ 
\\Department of Applied Mathematics and Theoretical Physics,\\The
University of Cambridge,\\Silver Street,\\Cambridge,\\CB3 9EW. \\ \\
{\bf damtp -- 1999 -- 89}}

\maketitle

\begin{abstract}
The thermodynamic properties of Schwarzschild-anti-de Sitter black
holes confined within finite isothermal cavities are examined. In
contrast to the Schwarzschild case, the infinite cavity limit may be 
taken which, if suitably stated, remains double valued. This allows
the correspondence between non-existence of negative modes for
classical solutions and local thermodynamic stability of the
equilibrium configuration of such solutions to be shown in a well
defined manner. This is not possible in the asymptotically flat
case. Furthermore, the non-existence of negative modes for the
larger black hole solution in Schwarzschild-anti-de Sitter provides
strong evidence in favour of the recent positive energy conjecture by 
Horowitz and Myers. 
\end{abstract}

\renewcommand{\thepage}{ }
\pagebreak
\renewcommand{\thepage}{\arabic{page}}
\setcounter{page}{1}

\sect{Introduction}
It is well known that if the free energy of a system has an imaginary
component, then that system is meta-stable and may decay via some
mechanism in a first-order phase transition. This effect is well
illustrated in, for example, the treatment of hot flat space in four 
dimensions given by Gross, Perry, and Yaffe~\cite{GPY}. In their
analysis, the authors demonstrate that flat space at zero temperature
is stable both classically and quantum-mechanically since the flat
space metric $\eta$, which is a true minimum of the action, is the
unique classical solution about which any saddle-point expansion may
be made. In the case of flat space at some non-zero temperature $T$
however, this is no longer true, since there exist other classical
solutions with period $\beta = 1/T$, such as the Euclidean section of
the Schwarzschild solution, which is not a true minimum of the action.  

The $n=4$ dimensions Euclidean Schwarzschild solution is seen to
possess a smooth -- i.e. monotonically decreasing in $r$ --
spherically symmetric metric perturbation which decreases the
Euclidean action. The existence of this unique nonconformal `negative
mode' demonstrates that the Schwarzschild solution is a saddle-point
of the action rather than a true minimum, and may be interpreted as a 
quantum-mechanical instability of hot flat space since nearby
geometries contribute an imaginary component to the free energy $F = -
1/\beta \: \log Z$ of the system. This permits, at any non-zero
temperature, the spontaneous excitation of hot flat space over an
effective potential barrier and the subsequent nucleation of black
holes with a rate proportional to the imaginary component. This
process may be viewed as a phase transition through topologies with
Euler characteristic $\chi = 0$ to $\chi = 2$, and is purely quantum
in nature with no classical analogue.     

The authors conclude, therefore, that asymptotically flat space at
any non-zero temperature $T$ is quantum mechanically unstable due to
the spontaneous nucleation of black holes with mass $M = 1 / 8 \pi
T$. This conclusion is essentially in agreement with the idea that the
canonical ensemble in hot flat space is pathological when gravitational
effects are taken into consideration, although in the form given by
Gross {\sl et al.} the nucleation process is understood to be
thermodynamically inconsistent.    

The above results pertain to black hole nucleation in hot flat space
confined to a large ideal isothermal cavity -- essentially in the
limit of infinite volume. However, a far deeper understanding of black
hole thermodynamics and the nucleation process is afforded by instead
considering cavities with finite volume~\cite{SWH1, GP1}. In this
approach, due to York~\cite{YORK1}, the canonical ensemble is
constructed from elements which are ideal isothermal boxes of
invariant surface area $A_B = 4 \pi r_B^2$ and wall temperature
$T_B$. The Euclidean Einstein equations with canonical boundary
conditions then admit either zero black hole solutions if the product
$r_B T_B < \sqrt{27}/8 \pi$, or else two distinct black hole solutions
which are degenerate at equality. The larger of the black hole
solutions is locally thermodynamically stable -- i.e. has positive
heat capacity, and may also have negative free energy, but it does not
have mass $1/8 \pi T_B$. The mass of the smaller black hole does
approach $1/8 \pi T_B$ in the limit of large $r_B$, but it is
thermodynamically unstable and has positive free energy.  

These observations allowed York to give a consistent interpretation of
the black hole nucleation process in a finite cavity. In a spontaneous
process such as that proposed the relevant thermodynamic potential --
the free energy in the canonical ensemble -- cannot increase, and so
it is the larger thermodynamically stable black hole which may form by
nucleation. The positive free energy of the smaller hole acts as the
effective potential barrier.

A further advantage of this approach to black hole thermodynamics is
that it demonstrates a clear equivalence between the requirements for
local thermodynamic stability of an equilibrium configuration in the
canonical ensemble and the requirements for local dynamic stability of
the classical solutions. This correspondence may be demonstrated using
the `reduced action' $I_{\star}$ derived by Whiting and
York~\cite{YORK2}. Whiting~\cite{WHITING1} shows explicitly that, at
equilibrium, the second derivative of the reduced action with respect
to the horizon radius is positive --  and so the action is a local
minimum rather than a saddle point --  only in the region where the
heat capacity is positive. This is when the cavity radius satisfies
the constraint $r_+ < r_B < 3 r_+ / 2$ where $r_+$ is the horizon
radius.   

Analysis of the Euclidean Schwarzschild nonconformal negative mode is
extended to isothermal cavities of finite volume by
Allen~\cite{ALLEN}, who finds that the mode persists until the cavity
radius $r_B$ falls below some critical radius $r_{\mr{crit}} \sim 2.89
\, r_+/2$. The non-existence of a negative mode for a classical
solution is a clear indication of local dynamic stability with respect
to small perturbations of the background, and provides the same
criterion on the action as the second derivative of $I_{\star}$ with
respect to the horizon radius. It should likewise correspond to the
sign of the heat capacity, but there is clearly some small amount of
disagreement here. This is at least partly due to Allen's choice of
boundary conditions: invariant cavity radius $r_B$ rather than
invariant surface area $A_B$, but as York has pointed
out~\cite{YORK1}, this implies that the eigenvalue spectrum of the
negative mode is not calculated at an extremum of the action. In fact,
the boundary conditions for calculating this spectrum depend
explicitly on the magnitude of the metric perturbation in the interior
region. It would appear, therefore, that in a finite isothermal cavity
there can be no well defined way of relating the local dynamics of a
classical solution in terms of the eigenvalue spectrum with the local
thermodynamics of the equilibrium configuration. Of course, the
original calculation of the negative mode eigenvalue in the infinite
volume case is properly defined: there is no choice in the boundary
conditions. However this demonstrates only that a dynamic instability
exists when the heat capacity is negative, not that the instability
vanishes when the solution becomes thermodynamically stable. In the
limit $r_B \rightarrow \infty$ the radius above which this occurs
becomes infinite, and the solution with positive heat capacity does
not exist. 

The characteristics of the canonical ensemble in hot flat space
confined to a finite cavity may be contrasted with those in the
presence of a negative cosmological constant. For this purpose, York's
thermodynamic analysis is extended to the case of asymptotically
anti-de Sitter black holes by Brown, Creighton and Mann~\cite{BCM},
who study their properties in both two and three spatial dimensions.
In the case of three spatial dimensions the results are qualitatively
similar to flat space: for a given invariant surface area $A_B$ there
will exist either zero or two black hole solutions to the Euclidean
Einstein equations with canonical boundary conditions. If the constant
boundary temperature $T_B$ is chosen such that two distinct solutions
exist, then the larger black hole will be thermodynamically stable and
the smaller black hole will be unstable.

A key difference between the two cases, however, lies in the behaviour
of the heat capacity. In Schwarzschild-anti-de Sitter this is positive
when the radius of the cavity satisfies the constraint $r_+ < r_B <
f(r_+)$, where $f(r_+) \rightarrow \infty$ as $r_+ \rightarrow b /
\sqrt{3}$ from below, with $f(r_+) = \infty$ for $r_+ > b /
\sqrt{3}$. This means that in the limit of the cavity becoming
infinite, and with an appropriate rescaling of the temperature on the 
boundary wall, there will still exist two well-defined black hole
solutions to the Euclidean Einstein equations: one locally
thermodynamically stable and the other unstable. In fact, if suitably
stated, all of the physical characteristics of the solutions are
retained in this limit, and the situation recovered is formally
identical to that which is analysed by Hawking and Page~\cite{PAGE}. 

In their approach, these authors identify the temperature fixed at
infinity with the inverse of the periodicity $\beta_{\star}$ in the
imaginary-time direction. The temperature then has a minimum value
$T_0$, below which there are no black hole solutions and above which
there are two. They conjecture that, like Schwarzschild, the
Schwarzschild-anti-de Sitter solution should have a single
nonconformal negative mode, but only when $r_+ < r_0$ where $r_0 = b /
\sqrt{3}$ -- the horizon radius of the unique black hole solution with
temperature $T_0$. They further conjecture that the negative mode
should `pass through zero' at $r_+ = r_0$, and that for solutions with
horizon radius greater than this, no nonconformal negative modes
should exist. This strongly suggests that the action of the larger
solution, should it be negative, is in fact a global minimum of the
action functional for asymptotically anti-de Sitter metrics
periodically identified in imaginary time.

The objective of this work is to confirm these conjectures by
explicitly calculating the eigenvalue spectrum for the nonconformal
negative mode in Schwarzschild-anti-de Sitter. Since the boundary of
the cavity is at infinity this calculation is inherently independent
of the magnitude of the metric perturbation. It will therefore provide
a well defined demonstration of the correspondence between the local
dynamic stability of a classical solution in terms of the existence of
a negative mode and the local thermodynamic stability of the
equilibrium configuration in the canonical ensemble. It would seem
that this is only possible when the cosmological constant is non-zero. 

The paper proceeds as follows: the next section is devoted to a review
of the thermodynamic properties of both asymptotically flat and
asymptotically anti-de Sitter black holes confined within a finite
ideal isothermal cavity. For simplicity the analysis is quite
graphical, and attention is restricted to the case of three spatial
dimensions only. Section three briefly discusses the theory behind
one-loop contributions to the partition function in the path-integral
approach to quantum gravity, and defines the physical gauge
independent operator $G$ acting on transverse and trace-free metric
perturbations. The eigenvalue equation for this operator acting on
spherically symmetric modes reduces to a second order O.D.E. in the
radial coordinate. In section four this equation is obtained in a
general form suitable for asymptotically flat and asymptotically
anti-de Sitter solutions in any dimension $n \ge 4$. Section five
discusses the numerical solution of this equation for
Schwarzschild-anti-de Sitter in $n=4$ dimensions and demonstrates the
stated correspondence with the heat capacity. This is then extended to
higher dimensions in section six. The final section discusses the
relevance of these observations to more recent work involving
anti-de Sitter spaces, and in particular to the positive energy
conjecture for periodically identified anti-de Sitter suggested by
Horowitz and Myers~\cite{MYERS}. The non-existence of a negative mode
for the thermodynamically stable Schwarzschild-anti-de Sitter black
hole provides strong evidence in favour of this proposal.

Units in which $G = \hbar = c = k_B = 1$ are used throughout the
paper.

\sect{Thermodynamic Properties of Black Holes in Finite
Isothermal Cavities}
In three spatial dimensions, the Euclidean Schwarzschild solution
takes the well known form
\be
ds^2 = V(r) \, d\tau^2 + \frac{1}{V(r)}\, dr^2 + r^2 \, d\Omega^2
\label{metric} 
\ee
where the metric function $V(r) = 1 - r_+ / r$, with $r_+ = 2
M$. Clearly this solution is positive-definite only for values of $r$
greater than $r_+$, and a coordinate singularity occurs in the
$(\tau,r)$ plane at $r = r_+$. The $\mr{S}^2$ at this point may be
included in the Euclidean section if, in the conventional manner, the
coordinate $\tau$ is identified periodically with a period of
\be
\beta_{\star} = \left. \frac{4 \pi}{V'(r)} \right|_{r = r_+} = \: 4
\pi r_+. 
\ee
Since the killing vector $\partial / \partial \tau$ is naturally
normalised to 1 in the limit of large $r$, the temperature measured at
infinity may be formally identified with the inverse of this
period. The Tolman law then states that for any static
self-gravitating system in thermal equilibrium, a local observer at
rest will measure a local temperature $T$ which scales as
$g_{00}^{-1/2}$. In the present context, then, the constant of
proportionality is $T_{\infty} = 1 / \beta_{\star}$.  

\begin{figure}
\begin{picture}(0,0)(0,0)
\put(84,36){\small $T_B \! \uparrow$}
\put(121,60){\small $T_B \! \downarrow$}
\put(200,240){\small $r_+ = r_B$}
\put(250,208){\small $r_+ = 8 r_B / 9$}
\put(230,140){\small $r_+ = 2 r_B / 3$}
\put(288,0){\small $r_B$}
\put(-9,265){\small $r_+$}
\put(32,165){\small Exterior Region}
\end{picture}   
\centering\epsfig{file=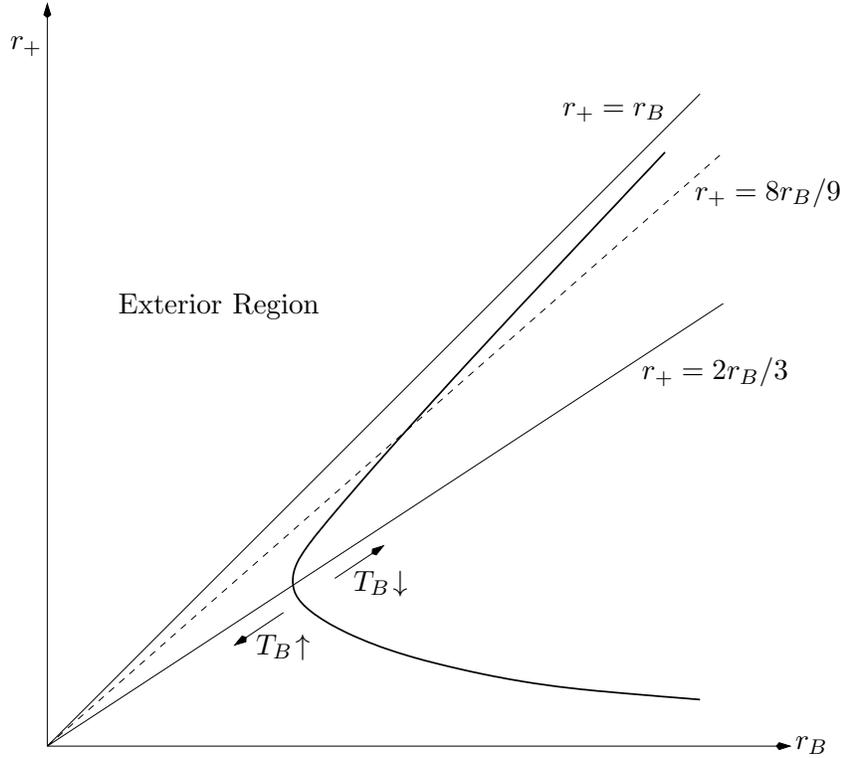,width=10cm}   
\caption{A constant temperature slice through the $(T_B,r_B,r_+)$
surface defined by (\ref{sch_surface}) for a Schwarzschild black hole
confined within an ideal isothermal cavity. Also shown are the lines
$r_+ = r_B$ and $2 r_B / 3$, between which the heat capacity $C_A$ is
positive. The broken line is $r_+ = 8 r_B / 9$, above which the free
energy is negative.}   
\label{sch}
\end{figure}

With this in mind, York~\cite{YORK1} defines the elements of the
canonical ensemble for hot flat space to be ideal isothermal cavities
of invariant surface area $A_B = 4 \pi r^2_B$ and wall temperature
$T_B$. One topologically regular solution to the Einstein equations
with these boundary conditions is hot flat space with a uniform
temperature of $T_B$ throughout the cavity. Another solution is the
Schwarzschild metric. If a black hole of horizon radius $r_+ < r_B$
does exist within the cavity then clearly, from the Tolman law, the
wall temperature must satisfy  
\be
T_B \definetobe T(r_B) = (4 \pi r_+)^{-1} \: (1 - r_+ / r_B)^{-1/2}.
\label{sch_surface}
\ee
This equation may be solved for $r_+$ in terms of the constants $r_B$
and $T_B$. If the product $r_B T_B < \sqrt{27}/8 \pi$ then there are
no real positive solutions for $r_+$. In this part of the
$(T_B,r_B)$ plane no black holes exist. If this inequality is not
satisfied then in general two distinct solutions exist, which join
smoothly at equality where $r_+ = 2 r_B / 3$. Figure~\ref{sch} shows
this curve as a constant temperature slice through the $(T_B,r_B,r_+)$
surface. 

The diagram is to be interpreted as follows: for some given wall
temperature $T_B$ the solution curve will appear as shown. If the
scale of $r_B$ along the horizontal axis is kept fixed, then for
higher wall temperatures the solution curve shifts to the left, and
for lower wall temperatures it shifts to the right. The turning point
of the curve, where $r_B T_B = \sqrt{27} / 8 \pi$, remains on the line
$r_+ = 2 r_B / 3$. It is clear therefore that for any cavity radius to
the right of this point there will be two distinct Schwarzschild black
hole solutions -- one above this line and one below it.

For any values of $r_B$ and $T_B$, the entropy of the black hole
solutions to (\ref{sch_surface}) is $S = \pi r_+^2$. The heat capacity
at constant surface area -- the analogue in this context of $C_V$ --
may therefore be calculated for any solution using 
\be
C_A \definetobe T_B \left. \frac{\partial S}{\partial T_B}
\right|_{A_B} = \: - 2 \pi r_+^2 \: (1 - r_+^{\p{2}}/ r_B^{\p{2}}) \:
(1 - 3 r_+^{\p{2}}/ 2 r_B^{\p{2}})^{-1}.
\ee
Hence, if $r_+ < r_B < 3 r_+ / 2$ the heat capacity is positive and
the equilibrium configuration is locally thermodynamically stable. As
figure~\ref{sch} shows, this is always the case for the larger black
hole solution.

A far deeper understanding of the thermodynamics of black holes
in finite cavities, and the associated roles of both classical and
non-classical geometries within the canonical ensemble, is afforded by
considering the `reduced action' $I_{\star}$ proposed by Whiting and 
York~\cite{YORK2, WHITING1, WHITING2, YORK3}. This is the function
obtained from the Euclidean Einstein action by considering only static
and spherically symmetric geometries which: i) are smooth and have
$\chi=2$ topology; ii) have appropriate boundary data for the
canonical ensemble; and iii) obey the constraints of the Einstein
equations on a family of space-like hyper-surfaces which foliate the
manifold. 

With these requirements the reduced action becomes
\be
I_{\star} = \beta_B \, E - S
\ee
where $\beta_B$ is the inverse wall temperature, $S$ is the classical
black hole entropy $\pi r_+^2$, and the energy is chosen to be $E =
r_B - r_B \, (1 - r_+ / r_B )^{1/2}$. With this definition, then, both
the energy and the action of hot flat space are zero.

It should be noted that the reduced action has the same form as the
action evaluated for a classical Schwarzschild solution, except that
the variables $\beta_B, r_B,$ and $r_+$ are in this case all
independent rather than being related through (\ref{sch_surface}).
With this function, therefore, it is possible to {\sl derive} both the
thermodynamic properties of the ensemble and the dynamic properties of
the classical solutions within the ensemble by varying with respect to
the single remaining degree of freedom $r_+$. Thus, for example, the
stationary points satisfying $\partial I_{\star} / \partial r_+ = 0$
yield $\beta_B = (4 \pi r_+) \: (1 - r_+ / r_B)^{1/2}$, in agreement
with the necessary classical requirement. Furthermore, at the
stationary points,  
\be
\frac{\partial^2 I_{\star}}{\partial r_+^2} = \frac{1}{4} \: (1 -
r^{\p{0}}_+ / r^{\p{0}}_B )^{-1} \: \beta_B^{-2} \: C^{\p{0}}_A,
\ee
and so 
\be
\frac{\partial^2 I_{\star}}{\partial r_+^2} > 0 \Leftrightarrow C_A
> 0. 
\ee
The condition for local dynamic stability of the classical solution is
therefore wholly equivalent to the condition for local thermodynamic
stability of the equilibrium configuration, where the latter is shown
to occur only within the range $r_+ < r_B < 3 r_+ / 2$. Should the
partition function in fact be dominated by a classical solution,
i.e. should the action for this solution -- and therefore the free
energy -- be negative, then the equilibrium configuration will be
globally thermodynamically stable. However, this occurs only within
the sub-range $r_+ < r_B < 9 r_+ / 8$. 

Following Whiting~\cite{WHITING1}, the preceding analysis may be
written in a gauge invariant form by instead varying the reduced
action with respect to a generalised function $\overline{\beta}(r_+)$.
This function is chosen such that $\partial S / \partial E \: |_{r_B}
= \overline{\beta}$, where the form of $E$ is left undefined.
Stationary points of $I_{\star}$ then occur at $\overline{\beta} =
\beta_B$, and the second variation with respect to $\overline{\beta}$
at the stationary points indicates local dynamic stability as before.

With the introduction of a negative cosmological constant $\Lambda$,
the characteristics of Euclidean solutions to the Einstein equations
are changed in a number of ways. The line element for these
solutions still takes the form (\ref{metric}), but the metric function
becomes $V(r) = 1 - 2M/r + r^2/b^2$ where $b^2 = - 3 / \Lambda$. This
is positive-definite only for values of $r$ greater than the horizon
radius $r_+$ -- the real positive root of $V(r)$, and the mass
parameter may be written in terms of this root as
\be
M = \frac{1}{2} \: r^{\p{0}}_+ \: ( 1 + r_+^2/b^2_{\p{0}} ).
\label{mass_func}
\ee
Again, there is a coordinate singularity in the $(\tau,r)$ plane at $r
= r_+$, and the $\mr{S}^2$ at this point may be included in the
Euclidean section if the period of $\tau$ is now fixed to be
\be
\beta_{\star} = \frac{4 \pi b^2 r_+}{b_{\p{0}}^2 + 3 r_+^2}.  
\ee
In this case, however, the temperature measured at infinity may no
longer be identified with the inverse of the periodicity as it can in
asymptotically flat spaces. There is no natural normalisation of the
Killing vector $\partial / \partial \tau$ in the limit of large $r$
due to the $r^2$ term in the metric function, and the locally measured
temperature of any thermal state decreases to zero at infinity.
Nonetheless, it is still consistent to regard $1 / \beta_{\star}$ as
the constant of proportionality in the Tolman equation. 

\begin{figure}
\begin{picture}(0,0)(0,0)
\put(87,24){\small $bT_B \! \uparrow$}
\put(119,60){\small $bT_B \! \downarrow$}
\put(200,240){\small $\rho_+ = \rho_B$}
\put(288,0){\small $\rho_B$}
\put(-9,265){\small $\rho_+$}
\put(265,52){\small $\rho_+ = C_0(\rho_B)$}
\put(265,91){\small $\rho_+ = F_0(\rho_B)$}
\put(-6,50){\small 1}
\put(-15,88){\small $\sqrt{3}$}
\put(32,165){\small Exterior Region}
\end{picture}   
\centering\epsfig{file=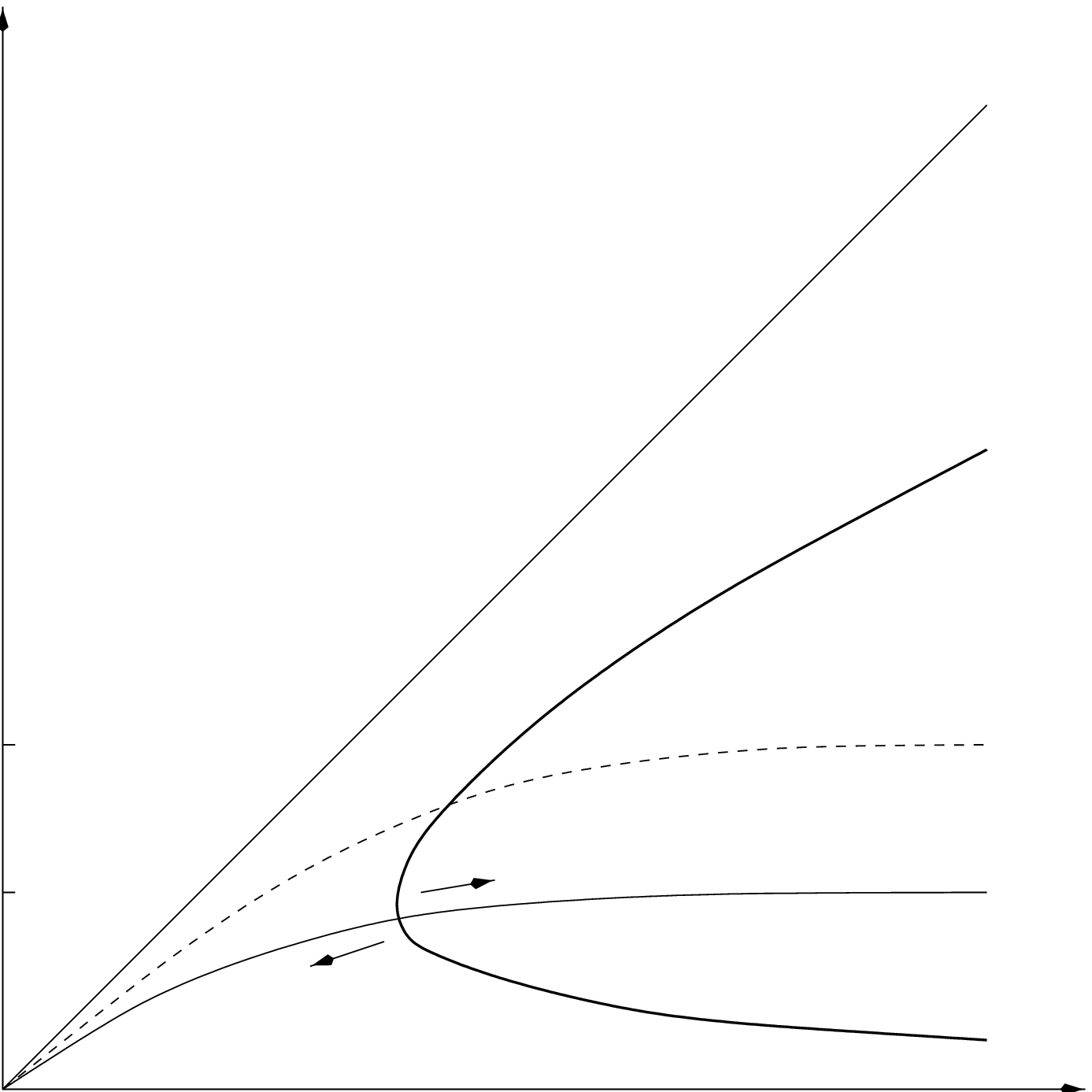,width=10cm}   
\caption{A constant temperature slice through the
$(bT_B,\rho_B,\rho_+)$ surface defined by (\ref{sch_ads_surface}) for
a Schwarzschild black hole in anti-de Sitter space confined within an 
ideal isothermal cavity. Also shown are the lines $\rho_+ = \rho_B$
and $C_0(\rho_B)$, between which the heat capacity $C_A$ is positive.
Note that $C_0(\rho_B) \rightarrow 1$ in the limit of large $\rho_B$. 
The broken line is $\rho_+ = F_0(\rho_B)$, above which the free energy
is negative.} 
\label{sch_ads}
\end{figure} 

So, the elements of the canonical ensemble for hot anti-de Sitter
space may again be defined as ideal isothermal cavities of invariant
surface area $A_B = 4 \pi r_B^2$ and wall temperature $T_B$, but
should there be a black hole within the cavity, this quantity must
satisfy 
\be
T_B \definetobe T(r_B) = \frac{b_{\p{0}}^2+3r_+^2}{4 \pi b^2 r_+} \:
r_B^{1/2} \: \bigl( r_B^{\p{0}} - r_+^{\p{0}} - r_+^3 / b^2_{\p{0}} +
r_B^3 / b^2_{\p{0}} \bigr)^{-1/2}.
\ee
Actually, it is convenient when considering anti-de Sitter to
transform to a dimensionless radial coordinate $\rho$, where
\be
\rho \definetobe \sqrt{3} \, r/b.
\label{rho_def}
\ee
Remaining factors of the constant $b$ may then be absorbed into the
definition of each quantity, although for clarity these are explicitly
written. The wall temperature then becomes  
\be
b T_B = \frac{3}{4 \pi} \: \frac{1 + \rho_+^2}{\rho_+} \: \rho_B^{1/2}
\: \bigl( \rho_B^3 + 3 \rho_B^{\p{0}} - \rho_+^3 - 3 \rho_+^{\p{)}} 
\bigr)^{-1/2}
\label{sch_ads_surface}
\ee
which may be solved for $\rho_+$ in terms of the constants $\rho_B$
and $bT_B$. Like the asymptotically flat case there will again exist
either zero or two distinct black hole solutions, governed by the
relative values of $\rho_B$ and $bT_B$, although the relation in this
case is not so clear. The two solutions are degenerate along the locus
$\rho_+ = C_0(\rho_B)$, defined implicitly through the real positive
root of 
\be
\rho_B^3 + 3\rho_B^{\p{3}} + \frac{1}{2} \: \left\{ \frac{ \rho_+^5 +
2 \rho_+^3 + 9 \rho_+^{\p{0}} }{ \rho_+^2 - 1} \right\} = 0.
\ee
Accordingly, for a given cavity radius, the temperature at and above
which solutions occur may be written as
\be
(bT_B)^2 = \frac{9}{8 \pi^2} \: \rho_B \: \frac{1-C_0^2}{C_0^3}.   
\ee
This behaviour is illustrated in figure~\ref{sch_ads}, which is to be
interpreted in the same way as the previous diagram for the
asymptotically flat case. It should be noted that the diagrams are
identical in the limit $\rho_B \ll 1$ -- i.e. close to the origin,
since the cosmological constant can have very little effect on the
physics within cavities of radius $r_B \ll b$. For larger cavities,
however, the differences are manifest.

The heat capacity at constant surface area is
\be
b^{-2} C_A = \frac{4 \pi}{3} \: \rho_+^2 \: \left\{ -1 + \frac{ 2
\bigl[ \rho_+^3 - 3 \rho_+^{\p{0}} \bigr] - (3 \rho_+^2 - 1) \bigl[
\rho_B^3 + 3 \rho_B^{\p{0}} \bigr] }{ (\rho_+^2+1) \bigl[ \rho_+^3 + 3
\rho_+^{\p{0}} - \rho_B^3 - 3 \rho_B^{\p{0}} \bigr] } \right\}^{-1}, 
\ee
which is negative below the line $\rho_+ = C_0(\rho_B)$ and positive
above it. As before, then, the equilibrium configuration of the larger
black hole solution is always locally thermodynamically stable, while
that for the smaller solution is always unstable. If the limit is
taken in which the cavity radius becomes large while the horizon radii
of the black hole solutions remain fixed~\cite{BCM}, then $b T_B
\rightarrow 0$ and 
\be
b^{-2} C_A \rightarrow \frac{2 \pi}{3} \: \rho_+^2 \: (\rho_+^2+1)
 / (\rho_+^2-1),
\label{C_A_limit}
\ee
clearly changing sign at $\rho_+ = 1$.

The reduced action approach to black hole thermodynamics may easily be
generalised to the $\Lambda \neq 0$ case. Choosing, for example, a
definition in which both the action and energy of hot anti-de Sitter
space vanish, the energy function becomes
\be
b^{-1} E = \frac{1}{3} \: \rho_B^{1/2} \: \left\{ \bigl( \rho_B^3 + 3
\rho_B^{\p{0}} \bigr)^{1/2} - \bigl( \rho_B^3 + 3 \rho_B^{\p{0}} -
\rho_+^3 - 3 \rho_+^{\p{0}} \bigr)^{1/2} \right\}.
\ee
Variation of the corresponding reduced action with respect to $\rho_+$
yields stationary points if the relation (\ref{sch_ads_surface}) is
satisfied, and the second variation at these points again demonstrates
local dynamic stability of the classical solution if and only if $C_A
> 0$. The free energy is negative only between $\rho_+ = \rho_B$ and
$\rho_+ = F_0(\rho_B)$, where $F_0$ is defined implicitly as the
solution curve for $\rho_+$ and $\rho_B$ along which the free energy
vanishes. Within this range the partition function will be dominated
by the classical solution, and so the equilibrium configuration will
be globally thermodynamically stable. Of course, Whiting's gauge
invariant proof of the correspondence between local dynamic and
thermodynamic stability is immediately applicable since it relies only
on the general form of $I_{\star}$, and not on the specific forms of
$E, \beta_B$ and $\overline{\beta}(\rho_+)$ individually.   

As the limit (\ref{C_A_limit}) makes clear, an important difference
between this case and the previous one lies in the asymptotic
behaviour of the functions $C_0$ and $F_0$. As the cavity size
increases, these tend to 1 and $\sqrt{3}$ respectively, in contrast to
their analogues in asymptotically flat space which both become
infinite. It is clear, therefore, that for an arbitrarily large cavity
radius  a sufficiently low wall temperature may always be chosen such
that both black hole solutions are of finite size. When the
cosmological constant is zero this is no longer the case since the
turning point of the solution curve, which lies along the line $r_+ =
2 r_B / 3$, occurs at an infinite radius.  

In the asymptotically flat case, then, the large black hole solution
is ill-defined as $r_B \rightarrow \infty$, and only the smaller
solution remains. The flat space canonical ensemble simply cannot
survive this limiting process. In the present case, however, with a
suitable rescaling of the wall temperature, a well-defined limit may
be taken in which the elements of the canonical ensemble are both
isothermal and infinite in extent, ostensibly because the functions
$C_0$ and $F_0$ remain finite. Furthermore, all of the physical and 
thermodynamic characteristics of the solutions in finite cavities are
retained. This latter observation is essential in the forthcoming
analysis.  

Since $\partial / \partial \tau$ possesses no natural normalisation at
infinity, a rescaling of the locally measured temperature merely
amounts to a rescaling of this Killing vector. The temperature is,
after all, only a Lagrange multiplier. With this in mind, if $T_B
\rightarrow \Delta \: T_B$ where the factor $\Delta$ is defined as  
\be
\Delta \definetobe \lim_{r \rightarrow \infty} \: V(r)^{1/2}
\ee
then, in the $\rho_B \rightarrow \infty$ limit, the fixed temperature
at infinity may be identified with $1 / \beta_{\star}$. The solution
recovered through this process is formally identical to that which is
analysed by Hawking and Page~\cite{PAGE}.

\begin{figure}
\begin{picture}(0,0)(0,0)
\put(290,0){\small $\rho_+$}
\put(2,240){\small $b^{-1}\beta_{\star}$}
\put(5,211){\small $\beta_{\mr{max}}$}
\put(68,5){\small 1}
\put(95,5){\small $\sqrt{3}$}
\put(-32,135){\small $C_A < 0$}
\put(160,130){\small $C_A > 0$}
\put(55,243){\small $F > 0$}
\put(100,243){\small $F < 0$}
\end{picture}   
\centering\epsfig{file=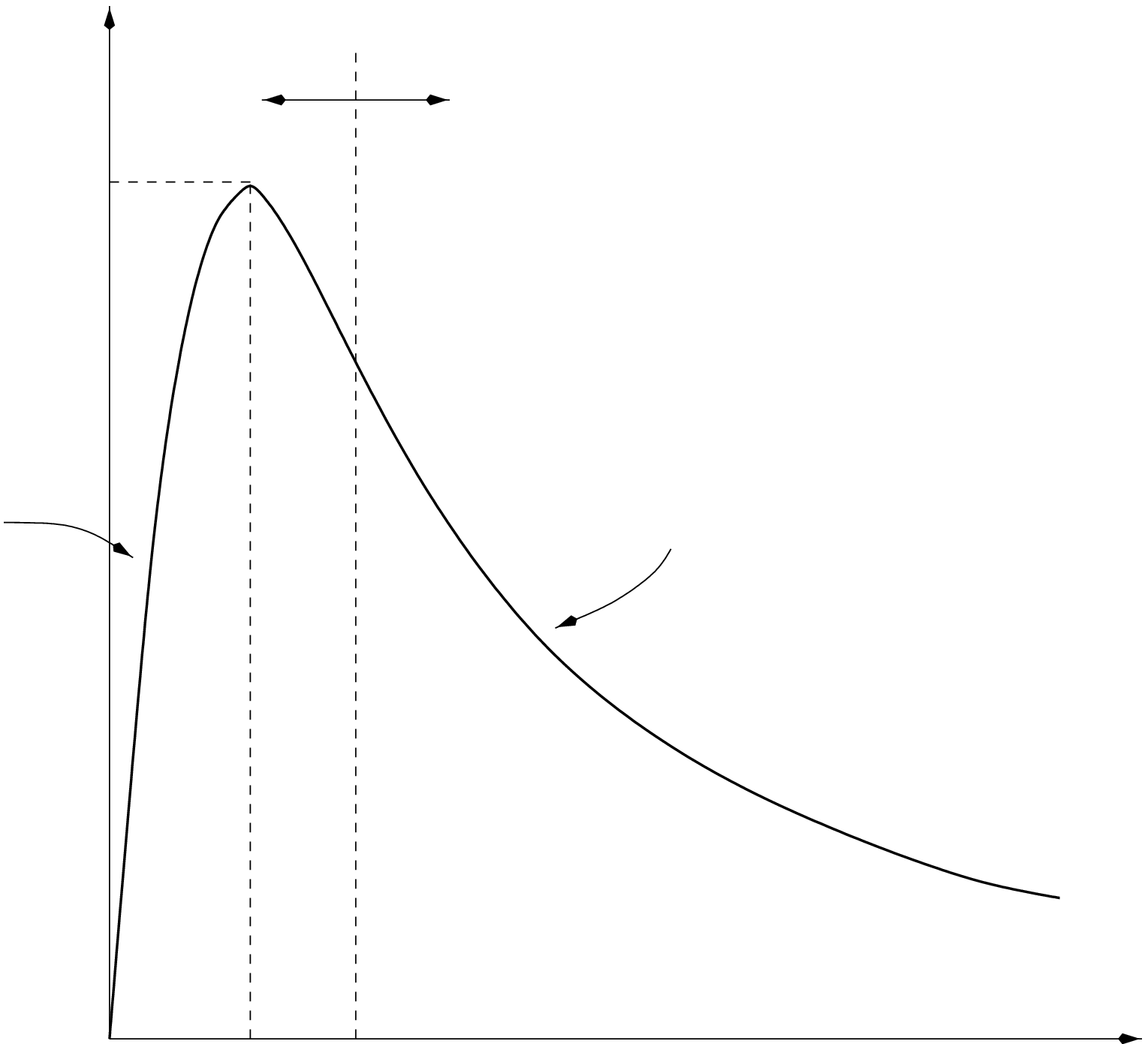,width=10cm}   
\caption{The black hole solution curve in the $(\rho_+,\beta_{\star})$
plane, in the limit of infinite isothermal cavity radius. The turning
point $\beta_{\mr{max}} = 1/T_0$ occurs at $\rho_+ = 1 \Leftrightarrow
r_+ = b / \sqrt{3}.$} 
\label{sch_ads_beta}
\end{figure} 

In the transformed coordinate $\rho$, the inverse temperature takes
the form
\be
b^{-1} \: \beta_{\star} = \frac{4 \pi}{\sqrt{3}} \: \rho_+^{\p{0}} /
(\rho_+^2 + 1).
\ee 
The features retained from the finite cavity case are immediately
apparent: in contrast to the infinite asymptotically flat case, a
minimum temperature still remains below which there are no black hole
solutions. This temperature occurs in the case of the unique solution
with horizon size $\rho_+ = 1$, where $b^{-1} \: \beta_{\mr{max}} = 2
\pi / \sqrt{3}$. For temperatures greater than $T_0 = 1 /
\beta_{\mr{max}}$ two distinct black hole solutions exist, with
horizon size larger and smaller than 1 respectively. As is manifest
from the limit (\ref{C_A_limit}), the heat capacity for the solutions
with $\rho_+ > 1$ is positive, while for those with $\rho_+ < 1$ it is
negative.

Again this may be related to the local dynamic stability using the
reduced action which, since $E=M$, takes the form
\be
I_{\star} = \frac{b}{2 \sqrt{3}} \: \beta^{\p{0}}_{\infty} \:
\rho_+^{\p{0}} \: (1 + \rho_+^2/3) - \frac{b^2}{3} \pi \rho_+^2.
\ee
Variation with respect to the horizon scale then yields
\be
\frac{\partial I_{\star}}{\partial \rho_+} = 0 \: \: \Leftrightarrow
\: \: \beta_{\infty} = \beta_{\star}
\ee
and 
\be
b^{-2} \: \left. \frac{\partial^2 I_{\star}}{\partial \rho_+^2} \:
\right|_{\beta_{\infty} =  \beta_{\star}} = \mbox{\hspace{5mm}}
\frac{2 \pi}{3} \: (\rho_+^2 - 1)/(\rho_+^2 + 1).
\ee
The sign of the second variation is then clearly the same as the sign
of the heat capacity for the equilibrium configuration. Furthermore,
by forcing the condition $\beta_{\infty} = \beta_{\star}$ in
$I_{\star}$, the action for the classical solutions becomes
\be
b^{-2} \: I = - \frac{\pi}{9} \: \rho_+^2 \: (\rho_+^2 - 3) /
(\rho_+^2 + 1), 
\ee 
from which it may be seen that the free energy indeed changes sign at
$\rho_+ = \sqrt{3}$, in agreement with the limiting value of
$F_0$. Clearly, both the action and the free energy are negative if
the horizon size is greater than $\sqrt{3}$.

These features of the infinite Schwarzschild-anti-de Sitter solution
are summarised in figure~\ref{sch_ads_beta}, and their significance
for the canonical ensemble is exhaustively explained by Hawking and 
Page~\cite{PAGE}. 

\sect{Nonconformal Negative Modes}
It is evident from the previous section, then, that the correspondence
between local thermodynamic stability of an equilibrium configuration
in the canonical ensemble and local dynamic stability of the classical
solutions is well established. This correspondence may be shown in a
gauge invariant way through variations of the reduced action for the
ensemble with respect to a generalised horizon radius, and is true for
both asymptotically flat and asymptotically anti-de Sitter space
confined within a finite ideal isothermal cavity. In the latter case,
however, both the canonical ensemble and the above correspondence
remain well defined, if suitably stated, even in the limit of infinite
cavity radius. 

Nonetheless, it is desirable to have a more direct and intuitive
indication of the dynamic stability of the classical solutions in
terms of a perturbation in the background geometry rather than in a
variation of the reduced action. The measure of stability in this
approach is the eigenvalue spectrum of the nonconformal perturbative
modes for the solution. Should there be a mode with a negative
eigenvalue, then the action for this solution is a saddle-point in its
phase space rather than a true minimum. Consequently, there ought to
be a correspondence between the presence of such a negative mode and
the local thermodynamic stability as governed by the heat capacity. As
will become clear in the forthcoming sections, it is the existence of
a formal limit in the anti-de Sitter case that permits a well defined  
prescription for this problem, which seems impossible in the
asymptotically flat case. 

Negative modes arise from the analysis of geometric fluctuations about
classical Euclidean solutions of the Einstein field equations.
However, the analysis to confirm their existence must be performed
with care, since the gauge freedom of the Euclidean action will in
general introduce a large number of non-physical negative modes
associated with conformal deformations of the metric. For pure
gravity, the contributions from the conformal and the nonconformal
modes decouple if a suitable gauge is chosen. This procedure is now
well understood. 

In the path integral approach~\cite{SWH2}, the partition function $Z$
is generally defined as a functional integral over all metrics with
some fixed asymptotic behaviour on some $n$-dimensional manifold
$\mc{M}$,   
\be
Z = \int_{\mc{M}} \! \! D[g] \: \mr{e}^{-i I[g]}.
\ee
This integral is formally defined by an analytic continuation to a
Euclidean section of $\mc{M}$ to become
\be
Z = \int_{\mc{M}} \! \! D[g] \: \mr{e}^{-\hat{I}[g]},
\label{partition}
\ee
where the integral is performed over all positive definite metrics
$g$. In the case of pure gravity, the Euclidean action $\hat{I}$ is 
\be
\hat{I} = -k \! \int_{\mc{M}} \! \! d^n x \: \sqrt{g} \: \left\{ R - 2
\Lambda \right\} - 2k \! \int_{\partial\mc{M}} \! \! d\Sigma \: \:
\mr{Tr} K 
\ee
where $k$ is a coupling constant and $K$ is the second fundamental
form on the boundary $\partial\mc{M}$.

This partition function may be approximated using saddle-point
techniques, by Taylor expanding about the known stationary points
of the Euclidean action -- the solutions to the Einstein field
equations 
\be
R_{ab} = \frac{2}{n-2} \Lambda \: g_{ab}.
\ee
The expansions are performed by writing the perturbed metric
$\tilde{g}$ as 
\be
\tilde{g}_{ab} = g_{ab} + \phi_{ab}
\ee
with $\phi$ treated as a quantum field on the classical fixed
background $g$ which vanishes on the boundary $\partial
\mc{M}$. Proceeding in this fashion yields 
\be
\hat{I}[\tilde{g}] = \hat{I}[g] + \hat{I}_2[\phi] + \cdots 
\ee
where the linear term $\hat{I}_1$ vanishes precisely because $g$
is a classical solution, $\hat{I}_2$ is quadratic in the field $\phi$,
and `$\cdots$' represents terms of higher than quadratic order.
Ignoring these additional terms, and inserting the remainder into
(\ref{partition}), gives the well-known expansion 
\be
\log Z = -\hat{I}[g] + \log \int_{\mc{M}} \! \! D[\phi] \: \mr{e}^{
-\hat{I}_2 [\phi]}.
\ee
The second term on the right is generally called the one-loop
contribution to $\log Z$.

The quadratic contribution to the action is straightforward to
evaluate, and may be written for arbitrary $\phi$ in the
form 
\be
\hat{I}_2[\phi] = \frac{k}{2} \int \! \! d^n x \: \sqrt{g} \:
\phi^{ab} A_{abcd} \: \phi^{cd}.
\ee
In this context, the operator $A$ takes the rather opaque
form~\cite{SWH2} 
\ba
\lefteqn{ A_{abcd} = \frac{1}{4} g_{cd} \nabla_a \nabla_b -
\frac{1}{4} g_{ac} \nabla_d \nabla_b + \frac{1}{8} \{ g_{ac} g_{bd} -
g_{ab} g_{cd} \} \nabla_e \nabla^e + \frac{1}{2} R_{ad} g_{bc} } & & 
\nonumber \\[2mm]
& & \mbox{\hspace{2cm}} -\frac{1}{4} R_{ab} g_{cd} + \frac{1}{16} R
g_{ab} g_{cd} - \frac{1}{8} R g_{ac} g_{bd} - \frac{1}{8} \Lambda
g_{ab} g_{cd} + \frac{1}{4} \Lambda g_{ac} g_{bd} \nonumber \\[3mm]
& & \mbox{\hspace{4.5cm}} + (a \leftrightarrow b) + (c \leftrightarrow
d) + (a \leftrightarrow b, c \leftrightarrow d).
\ea
Naively, then, the one-loop term may be written as $\frac{1}{2}
\log \mr{Det} (\mu^{-2} A)$ where $\mu$ is a regularisation mass, and
the determinant is formally defined as the product of the eigenvalues
of $A$. However, due to the diffeomorphism gauge freedom of the
action, $A$ will in general have a large number of zero eigenvalues,
and so this procedure as stated is ill-defined. The remedy is to add a
gauge fixing term $B$ -- such that the operator $A+B$ has no zero
eigenvalues -- and an associated ghost contribution $C$, to obtain
\be
\log Z = -\hat{I}[g] - \frac{1}{2} \log \mr{Det} (\mu^{-2} \{A+B\}) +
\log \mr{Det} (\mu^{-2} C).
\ee 

Such terms may be dealt with by means of generalised zeta functions,
as considered by Gibbons, Hawking, and Perry~\cite{GHP}, and extended 
to include a $\Lambda$ term by Hawking~\cite{SWH2}. In order that this
be possible, the terms must be expressed as sums of operators, each
with only a finite number of negative eigenvalues. This may be
achieved by writing $A+B$ as $-F+G$, where $F$ is a scalar operator
acting on the trace of $\phi$. The ghost term $C$ is a spin-1 operator
acting on divergence-free vectors, and $G$ is a physical gauge
invariant spin-2 operator which acts on the transverse and  
trace-free part of $\phi$. For any dimension $n \ge 4$, $G$ takes the
simple form 
\be
G_{abcd} = -g_{ac}g_{bd} \nabla_e \nabla^e - 2 R_{acbd}.
\ee

An observation of primary significance is that, for $\Lambda \le 0$, a
gauge may be chosen in which the operators $F$ and $C$ have no
negative eigenvalues. If the background metric $g$ is flat then, in
addition, $G$ will be positive-definite, but for a non-flat background
this is not the case. In general, $G$ will have some finite number --
generally zero or one -- of negative eigenvalues, which correspond to
the nonconformal negative modes of the solution. The eigenvalues of
$G$ are determined by all solutions to the elliptic equation 
\be
G^{ab}_{\p{ab}cd} \: \phi_{(n)}^{\p{n}cd} = \lambda^{\p{a}}_{(n)} \: 
\phi_{(n)}^{\p{n}ab} 
\label{G_eq}
\ee
where the eigenfunctions $\phi$ are real, regular, symmetric,
transverse, trace-free, and normalisable tensors. Clearly, should one
of the eigenvalues of $G$ be negative, then the product of all of the
eigenvalues would also be negative. The contribution to $\log Z$ from
fluctuations about the classical solution would then contain an
imaginary component, leading to an instability in the ensemble similar
to the type proposed by Gross, Perry and Yaffe~\cite{GPY}. 

\sect{General Spherically Symmetric Solutions}
Following the treatment given in~\cite{GPY}, and subsequently
in~\cite{ALLEN}, for the Schwarzschild case, it is clear that only
spherically symmetric and $\tau$-independent solutions of (\ref{G_eq})
need be considered as candidate nonconformal negative modes. In static
and spherically symmetric backgrounds, modes of higher multipole
moment will necessarily have greater eigenvalues. With this
assumption, it is then straightforward to write down a construction
for such solutions to $G$ valid in {\sl any} $n$-dimensional Euclidean 
black hole background of the form (\ref{metric}) for a general metric
function $V(r)$. In the following expressions, a prime denotes $d/dr$. 

Since $G$ acts only on symmetric transverse and trace-free tensors,
then clearly the constructed solutions must exhibit all of these
properties. If the mode $\phi_{ab}$ is written in the manifestly
trace-free and symmetric form 
\be
\phi^a_{\p{a}b} = \mr{diag} \: \Big( \psi(r), \chi(r),
\underbrace{k(r), \ldots, k(r)}_{(n-2) \: \mr{terms}} \Big)
\label{tracefree}
\ee
where the function
\be
k(r) = \frac{1}{2-n} \{ \psi(r) + \chi(r) \},
\ee
then the final property, $\nabla^a \phi_{ab}$ = 0, is guaranteed if it
is further assumed that $\psi(r)$ and $\chi(r)$ are related through
the first order equation
\be
\psi(r) = \frac{2rV}{rV'-2V} \: \chi'(r) +  \frac{rV'+2(n-1)V}{rV'-2V}
\: \chi(r).
\label{transverse}
\ee

With the ansatz (\ref{tracefree}) and (\ref{transverse}), the
eigenvalue equation (\ref{G_eq}) reduces to a linear second order
ordinary differential equation for the component $\chi(r)$ which, for
general $V(r)$, is
\ba
\lefteqn{ -V \chi''(r) +  \left[ \frac{2r^2 [ VV''- V'^2 ] -
r(n-2)VV'+ 2nV^2}{r(rV'-2V)} \right] \chi'(r) } & & \nonumber \\[3mm]
 & & \mbox{\hspace{15mm}} + \left[ \frac{r^2V'V'' + r [ 2(n-1)VV'' -
(n+2)V'^2 ] +4VV'}{r(rV'-2V)} \right] \chi(r) = \lambda \chi(r).
\label{diff_eq}
\ea
The problem of finding the eigenfunctions of $G$ is therefore reduced,
in the cases of interest, to the problem of finding real and
normalisable solutions to this equation that satisfy appropriate
boundary conditions.

For the backgrounds of interest, (\ref{diff_eq}) may have as many as
five singular points, only three of which are relevant. These are at
the horizon radius $r = r_+$, at $r = r_s$ -- the real solution to
$rV' - 2V = 0$, and at $r = \infty$. It is clear from
(\ref{transverse}) that at the first of these, where $V(r)$ vanishes,
the components $\psi(r)$ and $\chi(r)$ are equal and so the metric
perturbation does not alter the periodicity of $\tau$. Furthermore, in
order that $\phi_{ab}$ be  regular everywhere, the product $(rV'-2V)
\: \psi(r)$ must vanish at the second singular point, and hence 
\be
\chi'(r_s) / \chi(r_s) \: = \left. - \frac{1}{2rV} \: \biggl[rV' +
2(n-1)V \biggr] \: \right|_{r = r_s}.  
\label{f_deriv}
\ee

The normalisation of the eigenfunctions may be defined as  
\be
\mc{N}^2 = \int_{\mc{M}} \!\! d^nx \: \sqrt{g} \: \phi^{ab} \,
\phi_{ab},
\label{big_norm}
\ee
so that acceptable solutions must be square-integrable functions in
the sense that $\mc{N}^2$ be finite, although the extent of $\mc{M}$
will depend on the situation under consideration. (Henceforth, any
reference to the solution will imply the normalised function
$\mc{N}^{-1} \phi_{ab}$, unless $\mc{N}$ is explicitly written.) In
the infinite cavity limit this requirement, and the condition of
regularity at the interior singular points, provide the necessary
boundary conditions to solve for viable solutions. For black holes
confined within finite cavities however, this is no longer the case,
since all regular solutions to (\ref{diff_eq}) will be normalisable.
In such cases the canonical boundary conditions for the ensemble must
be imposed at the cavity wall, and it is this requirement which
presents difficulties. 

The perturbed metric $g_{ab} + \epsilon \phi_{ab}$ induces a line
element 
\be
ds^2  = V(r) \biggl[ 1 + \epsilon \psi(r) \biggr] \: d \tau^2 +
\frac{1}{V(r)} \biggl[ 1 + \epsilon \chi(r) \biggr] \: dr^2 + r^2 
\biggl[1 + \epsilon k(r)\biggr] \: d \Omega^2,   
\ee
and so the proper length around the $\mr{S}^1$ in the $\tau$ direction
at radius $r_B$, which may be identified with $1 / T_B$, becomes 
\be
1 / T_B = \beta_{\star} \: V(r_B)^{1/2} \: \bigg[1 + \epsilon
\psi(r_B) \biggr]^{1/2}. 
\ee
For an isothermal cavity wall at $r = r_B$ the temperature must
remain constant and so the boundary condition, as used by
Allen~\cite{ALLEN}, is 
\be
\psi(r) \: |_{r = r_B} = 0. 
\label{bc1}
\ee
However, this condition is insufficient, since the elements of the
canonical ensemble are defined by both an invariant wall temperature
and an invariant wall area $A_B$. In the perturbed metric the surface
area of a space-like spherical shell at radius $r_B$ is no longer $4
\pi r_B^2$, but becomes  
\be
\tilde{A}_B = 4 \pi r_B^2 \: \biggl[1 + \epsilon k(r_B) \biggr].
\label{new_area}
\ee
As York points out~\cite{YORK1}, evaluating the eigenvalue spectrum
with the boundary condition (\ref{bc1}) is not physically meaningful
for the classical solutions since the wall area at constant radius is
not invariant. This explains the discrepancy between the cavity radius
$r_B = 3 r_+ / 2$ at which the heat capacity of a Schwarzschild black
hole changes sign, and the critical radius at which the negative mode
vanishes in this case, calculated by Allen to be $r_{\mr{crit}} \sim
2.89 \, r_+ / 2$.  

So, to apply the correct boundary condition, a modified cavity radius
$r_{B'}$ must be chosen such that $\tilde{A}_{B'} = A_B$, and then the
condition 
\be
1 + \epsilon \psi(r_{B'}) = V(r_B) / V(r_{B'}) 
\ee
must be imposed. However, it is clear from (\ref{new_area}) that,
other than in the degenerate case $k(r) = \psi(r) = \chi(r) \equiv 0$,
this procedure is not particularly well defined since the parameter 
$\epsilon$ is freely variable. The areas $A_B$ and $\tilde{A}_{B'}$
may therefore be matched at any nearby radius simply by adjusting
$\epsilon$ appropriately. Then for a given choice of $\epsilon$ the
modified cavity radius will be uniquely defined, and the eigenvalue
spectrum of (\ref{diff_eq}) may be evaluated, but this spectrum will
in general be slightly different for each value of $\epsilon$.   

This problem disappears when the cavity is infinite. If the solution
$\phi_{ab}$ is normalisable in the sense discussed above, then the
radial functions $\psi(r)$, $\chi(r)$ and $k(r)$ must tend to zero at
some appropriate rate governed by the asymptotic form of
(\ref{diff_eq}). Consequently, the isothermal boundary condition at
infinity is automatically satisfied.  

The original calculation of the negative mode eigenvalue for
Schwarzschild in $n = 4$ dimensions, which is manifestly invariant of
the magnitude of $\epsilon$, finds $M^2 \: \lambda_{\mr{neg}} \sim
-0.192$. In the case of Schwarzschild-anti-de Sitter, a similar
calculation will achieve two goals. It will confirm the conjecture
made by Hawking and Page regarding the existence of a nonconformal
negative mode and its vanishing point at the critical radius $\rho_+ =
1$. It will also provide a well defined demonstration of the
correspondence between the local dynamic stability of a solution in
terms of its eigenvalue spectrum and the local thermodynamic stability
of the equilibrium configuration. This is only possible since, in
contrast to the flat case, the infinite cavity solution in the
asymptotically anti-de Sitter case retains all of the physical and
thermodynamic characteristics of the finite cavity solutions.      

\sect{The Eigenvalue Spectrum in Four Dimensions}
For Schwarzschild-anti-de Sitter in $n=4$ dimensions, the metric
function is
\be
V(r) = \frac{1}{r} \: \biggl( r - r_+^{\p{0}} - r_+^3/b^2_{\p{0}} +
r^3_{\p{0}}/b^2_{\p{0}} \biggr),
\ee
in which the horizon radius $r_+$ has been substituted for the mass
function $M$ from (\ref{mass_func}). Moving to the dimensionless radial
coordinate $\rho$ defined in (\ref{rho_def}), and defining a new
parameter $\alpha$ such that
\be
\alpha \definetobe \frac{1}{4} \: \bigl( \rho_+^3 + 3 \rho_+^{\p{0}}
\bigr),
\ee
the second order equation for $\chi(\rho)$ may be written in the
homogeneous form
\ba
\lefteqn{ \biggl[ \rho \, ( \rho-2\alpha) \, (\rho^3+3\rho-4\alpha)
\biggr] \: \chi''(\rho) + 4 \biggl[ 2 \rho^4 - 5 \alpha \rho^3 + 3
\rho^2 - 11 \alpha \rho + 8 \alpha^2 \biggr] \: \chi'(\rho) } & &
\nonumber \\[2mm] 
 & & \mbox{\hspace{5cm}} + \biggl[ (10 + \tilde{\lambda}) \rho^3 - 2
\alpha ( 18 + \tilde{\lambda}) \rho^2 - 16 \alpha \biggr] \:
\chi(\rho) = 0,
\label{4d_diff_eq}
\ea 
where a prime now denotes $d/d\rho$, and $\tilde{\lambda} = b^2
\lambda$. 

The three real singular points of this equation are at $\rho =
\rho_+$, $2 \alpha$, and $\infty$. The remaining two imaginary
singular points are irrelevant in this analysis. Around $\rho =
\rho_+$, a trial solution of the form $\chi(\rho) = \sum_n \: a_n \:
(\rho - \rho_+)^{k+n}$ yields the indicial roots $k = 0$ and $-1$. The
$k = 0$ case gives the unique regular series since the $k = 1$ series
is not normalisable in the sense that $\mc{N}^2$ is infinite. If $a_0$
is chosen to be 1, implying the boundary condition $\chi(\rho) \:
|_{\rho = \rho_+} = 1$, then the next two coefficients become  
\be
a_1 = - \frac{1}{6} \frac{ \rho_+^2 (18 + \tilde{\lambda}) +
24}{\rho_+^{\p{0}} (\rho_+^2 + 1)} 
\ee
and 
\be
a_2 = \frac{1}{108} \frac{ \rho_+^4 ( 648 + 54 \tilde{\lambda} +
\tilde{\lambda}^2_{\p{0}}) + \rho_+^2 (1620 + 66 \tilde{\lambda}) +
1008}{ \rho_+^2 ( \rho_+^2 + 1)^2_{\p{0}}}.
\ee
A similar expansion with coefficients $b_n$ around $\rho = 2 \alpha$
gives $k = 0$ and 3, both of which produce regular solutions. However,
only the $k = 0$ case contributes terms of order less than
$(\rho-2\alpha)^3$, for which the first three terms are related through
\be
b_1 = - \frac{2}{\alpha} \: b_0
\ee
from (\ref{f_deriv}), and 
\be
b_2 = \frac{1}{2} \frac{ \alpha^2 (18 + \tilde{\lambda}) + 6}{\alpha^2
(4 \alpha^2+1)} \: b_0.
\ee 
Using these coefficients a numerical solution to (\ref{4d_diff_eq})
may be generated which is constrained to within $\mc{O}(\epsilon^3)$
for $\epsilon \ll 1$ at the interior singular points. Any simple
algorithm is limited to this accuracy by the existence of the $k=3$
series about $\rho = 2 \alpha$.

For a given choice of horizon size $\rho_+$, the aim of such a
numerical method is to find the spectrum of eigenvalues
$\tilde{\lambda}_{(n)}$, and in particular the minimum eigenvalue
$\tilde{\lambda}_{\mr{neg}}$, for which regular and normalisable
solutions $\chi(\rho)$ exist. Having fixed $\rho_+$, then, a value for
$\tilde{\lambda}$ is guessed, and an appropriate numerical integrating
routine~\cite{NUM_REC} is used to integrate away from the horizon
using the coefficients $a_n$ as initial data. The solution is
calculated to within a small step $\delta$ of $2 \alpha$, and
then analytically continued through the singular point to $2 \alpha +
\delta$ using the $b_n$. From here, the solution is then calculated
up to some large radius, at which point the criterion of
normalisability is assessed. For most choices of $\tilde{\lambda}$
the solution will fail, but for some choices -- specifically the
eigenvalues -- it will not. 

The conditions upon which normalisability is assessed may be obtained
from the specific form in this case of $\mc{N}^2$ and the asymptotic
behaviour of (\ref{4d_diff_eq}). With the benefit of relations
(\ref{tracefree}) and (\ref{transverse}), the relation
(\ref{big_norm}) may be translated into an integral condition in one
dimension on $\chi(\rho)$, so that
\be
\mc{N}^2 = 4 \pi \beta^{\p{0}}_{\star} \: \int_{\rho_+}^{\infty} \! \!
d \rho \: \Theta(\rho) \: \chi^2(\rho). 
\ee 
The function $\Theta(\rho)$ is well defined and finite at the interior
singular points, and has asymptotic behaviour $\sim \rho^6$ as $\rho
\rightarrow \infty$. This behaviour is, of course, the same as that
for the weight function derived from the self-adjoint form of
(\ref{4d_diff_eq}). The solution $\phi_{ab}$ will therefore be
normalisable if $\chi(\rho) \sim \rho^{-(7/2+x)}$ for any
positive $x$.

\begin{figure}
\centering\rotatebox{270}{\epsfig{file=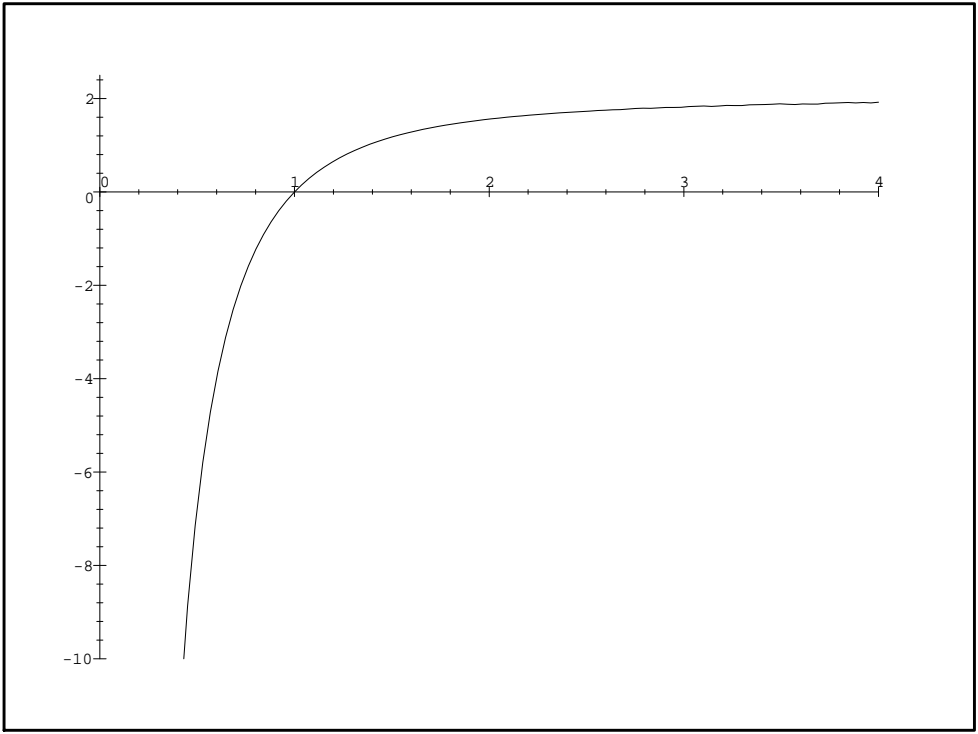,width=11cm}}   
\caption{Numerically generated results for
$\tilde{\lambda}_{\mr{neg}}$ (vertical) against $\rho_+$ in $n=4$
dimensions Schwarzschild-anti-de Sitter.}
\label{4d_ads}
\end{figure} 

The asymptotic form of (\ref{4d_diff_eq}) is readily obtained with the
standard substitution $\rho \rightarrow 1/u$. In the limit $u
\rightarrow 0$ this equation becomes
\be
u^2 \: \ddot{\chi}(u) - 6u \: \dot{\chi}(u) + [10 + \tilde{\lambda}]
\: \chi(u) = 0
\label{asymptotic_eq}
\ee
where an over-dot denotes $d/du$ -- the solutions to which describe the
asymptotic behaviour of $\chi(\rho)$. It should be noted that, in
contrast with the asymptotically flat case, the singular point at
$\rho = \infty$ is now regular. In this limit, 
\be
\chi(\rho) \sim A(\tilde{\lambda}) \: \rho^{-7/2 + \sqrt{9/4 -
\tilde{\lambda}}} \: + \: B(\tilde{\lambda}) \: \rho^{-7/2 - \sqrt{9/4
- \tilde{\lambda}}}. 
\label{asymptotic_sol}
\ee

Two points are immediately apparent from the form of this solution.
Firstly, in accordance with the naive expectation that the
eigenfunction corresponding to the lowest eigenvalue is the
`smoothest', the value of $\tilde{\lambda}_{\mr{neg}}$ ought to be
strictly less than 9/4. For eigenvalues greater than this the exponent
term acquires an imaginary component, and the solutions oscillate in
$\log \! \rho$. Secondly, the criterion of normalisability is clearly
equivalent to the single requirement $A(\tilde{\lambda}) \equiv 0$.

A simple numerical algorithm for finding $\tilde{\lambda}_{\mr{neg}}$
is therefore to proceed as described above and, for some fixed
$\rho_+$ and a guessed $\tilde{\lambda}$, integrate $\chi(\rho)$ out
to the second interior singular point at $2 \alpha$. Then, transform
the solution to a new function $\xi(\rho) = \rho^{7/2} \: \chi(\rho)$
and integrate this out to some large radius. As $\rho$ increases the
new function approaches  
\be
\xi(\rho) \sim A(\tilde{\lambda}) \: \rho^{\sqrt{9/4 -
\tilde{\lambda}}}
\ee
since the second term must be very small. For some choice of
$\tilde{\lambda}$, then, this remaining term will be negative while
for another it will be positive -- the two choices bracketing the true
value at which $A(\tilde{\lambda})$ vanishes. By subsequently refining
the range over which $\tilde{\lambda}$ is guessed the minimum
eigenvalue can be found to reasonable accuracy. This algorithm is
inelegant but nonetheless quite effective. Results for $\rho_+$ in
the range $0.1 \rightarrow 4.0$ are drawn in figure~\ref{4d_ads}.
    
The graph clearly shows a single negative mode for $n=4$ dimensions
Schwarzschild-anti-de Sitter vanishing at $\rho_+ = 1$. This confirms
the conjecture of Hawking and Page~\cite{PAGE}, and demonstrates in a
well defined manner the correspondence between this eigenvalue
spectrum and the local thermodynamic stability of the equilibrium
configuration in the canonical ensemble. As an aside, it should be
noted that 
\be
\lim_{\rho_+ \rightarrow 0} \: M^2 \: \lambda_{\mr{neg}} \sim 
-0.192
\ee
which demonstrates that the results obtained previously for
Schwarzschild reappear quite naturally in the limit where $r_+ \ll b$.

\sect{Extending the Spectrum to Higher Dimensions}
The various quantities employed in the evaluation of the eigenvalue
spectrum in $n=4$ dimensions Schwarzschild-anti-de Sitter are readily
generalised to any higher dimension. The form of the master equation
(\ref{diff_eq}) then makes it simple to verify that a nonconformal
negative mode will exist for a classical solution in any dimension $n
\ge 4$ only when the heat capacity of the equilibrium configuration is
negative.

In $n$ dimensions, the metric function $V(r)$ for Schwarzschild-anti-de
Sitter becomes
\be
V(r) = 1 - 2M / r^{n-3} + r^2/b^2
\ee
where $b^2 = -(n-1)(n-2)/2 \Lambda$ and so, if the horizon is at $r_+$
such that $V(r_+) = 0$, the mass may be expressed as
\be
M = \frac{1}{2} \: r_+^{n-3} \: (1 + r_+^2 / b^2_{\p{0}}).
\ee
The periodicity then generalises to
\be
\beta_{\star} = \frac{4 \pi b^2 r_+}{(n-3)b^2_{\p{0}} + (n-1)r_+^2}
\ee
and so the heat capacity becomes
\be
C_A = 2 \pi r_+^{n-2} \: \left[ \frac{ (n-1)r_+^2 + (n-3)b^2_{\p{0}}
}{ (n-1)r_+^2 - (n-3)b^2_{\p{0}} } \right].
\ee
The form of the heat capacity suggests the definition of a
dimensionless radial coordinate $\rho$ such that
\be
\rho \definetobe (n-1)^{1/2} \: (n-3)^{-1/2} \: r/b
\ee 
which clearly reduces to (\ref{rho_def}) when $n=4$. In any dimension
then, there is an infinite discontinuity in $C_A$ at $\rho_+ = 1$,
with $C_A$ negative for $\rho_+ < 1$ and positive for $\rho_+ > 1$. 

Inserting into (\ref{diff_eq}) any $n$, and with it the appropriate
form of $V(r)$ obtained from above, yields a second order equation for
$\chi(\rho)$ analogous to (\ref{4d_diff_eq}). Two of the three
relevant singular points of this equation remain fixed -- at $\rho_+$
and $\infty$, while the third is dimension-specific. All three remain
regular, however, regardless of the dimension. 

The weight function obtained from the self-adjoint form has 
asymptotic behaviour $\sim \rho^{n+2}$, while the general solution to
the asymptotic form, obtained with the substitution $\rho \rightarrow
1/u$, is
\be
\chi(\rho) \sim A(\tilde{\lambda}) \: \rho^{-(n+3)/2 + \beta} +
B(\tilde{\lambda}) \: \rho^{-(n+3)/2 - \beta}
\ee
where
\be
\beta^2 + \tilde{\lambda} = (n-1)^2 / 4.
\ee
The criterion of normalisability clearly remains the same --
i.e. $A(\tilde{\lambda}) \equiv 0$, and $\tilde{\lambda}_{\mr{neg}} <
(n-1)^2 / 4$.  

\begin{figure}
\centering\rotatebox{270}{\epsfig{file=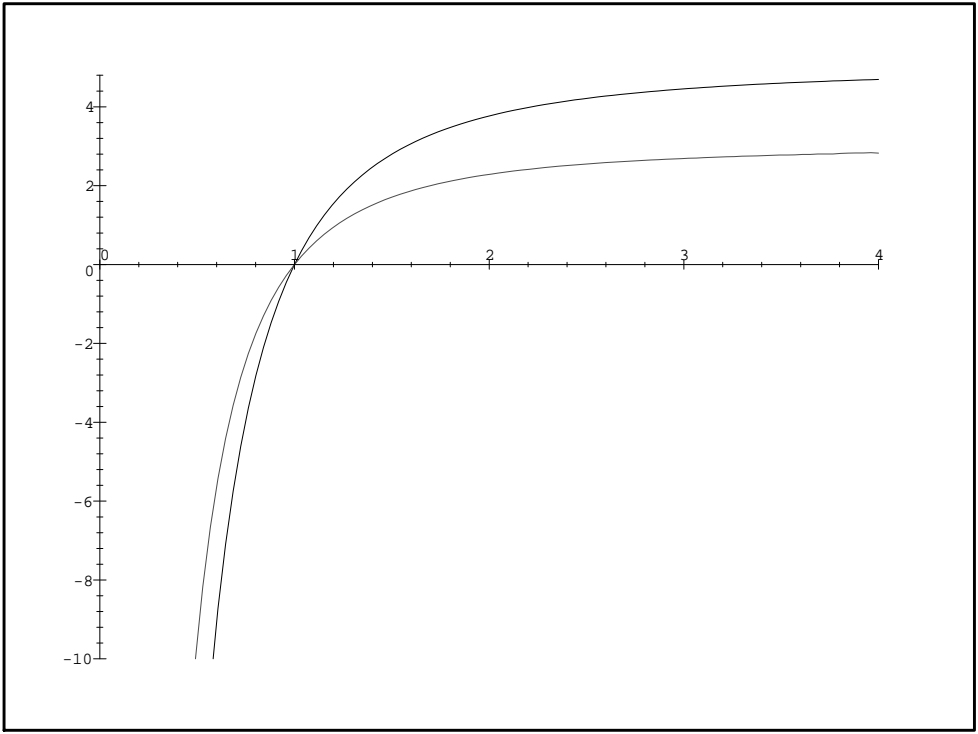,width=11cm}}   
\caption{Numerically generated results for
$\tilde{\lambda}_{\mr{neg}}$ (vertical) against $\rho_+$ in $n=5$ and
$n=7$ dimensions Schwarzschild-anti-de Sitter. The spectrum for $n=7$
gives the `upper' curve (tending to 5), and for $n=5$ the `lower'
curve (tending to 3).}
\label{5d_and_7d_ads}
\end{figure}

The numerical results for $\tilde{\lambda}_{\mr{neg}}$ obtained for
the cases $n=5$ and 7 for $\rho_+$ in the range $0.1 \rightarrow 4.0$
are shown together in figure~\ref{5d_and_7d_ads}. The conjecture of
Hawking and Page, and the correspondence between this eigenvalue
spectrum and local thermodynamic stability, may clearly be extended
to dimensions greater than four.

\sect{Discussion and Further Applications}
The discovery of a unique nonconformal negative mode for the $n=4$
dimensions Euclidean Schwarzschild solution led Gross, Perry and Yaffe
to propose the instability of flat space at finite temperature due to
the barrier-tunnelling nucleation of black holes. The tunnelling process
is permitted because, in the semi-classical approach, an integration
over quadratic fluctuations of the Schwarzschild solution produces an 
imaginary component in the free energy of the system. The existence of
a negative mode is therefore a powerful indication of instability.

The subsequent observation that the Euclidean Schwarzschild solution
becomes double-valued when confined within a box, and that hot flat
space may therefore be stabilised against black hole nucleation,
enabled York to give a thermodynamically consistent interpretation of
the tunnelling process within a finite isothermal cavity. Manifest
within this approach is a correspondence between local dynamic
stability of the classical black hole solutions and local
thermodynamic stability of the equilibrium configurations in the
canonical ensemble. This becomes a global thermodynamic stability if
the action, and so the free energy, is negative. It should also be the
case, therefore, that this correspondence applies to negative modes: a
nonconformal negative mode should exist only in those classical
solutions for which the heat capacity of the equilibrium configuration
is negative. However, there appears to be no well defined way of
demonstrating this explicitly in finite isothermal cavities. 
 
This problem may be overcome by instead considering hot anti-de Sitter
space. Confined again to a finite isothermal cavity, the Euclidean
Schwarzschild anti-de Sitter solution is also double-valued, but in
this case it remains so, if suitably stated, even in the infinite
cavity limit. The result so obtained is formally identical to that
analysed by Hawking and Page, and retains all of the physical and
thermodynamic characteristics of the finite cavity solutions. This
observation not only implies that the canonical ensemble in anti-de
Sitter is well defined, but also presents an ideal opportunity in
which to demonstrate the expected correspondence between negative
modes and heat capacity.  

Restricting the master equation (\ref{diff_eq}) derived above to the
case of $n=4$ dimensions Schwarzschild-anti-de Sitter, and solving for
the eigenvalue $\tilde{\lambda}_{\mr{neg}}$ of the smoothest
normalisable eigenfunction, results in an eigenvalue spectrum for a
range of classical solutions which indeed has the same sign as the
heat capacity function. This demonstrates the stated correspondence in
a manner invariant of the magnitude of the metric perturbation, and
confirms the conjecture of Hawking and Page that the negative mode
`passes through zero' at the solution with critical horizon radius
$r_+ = b/ \sqrt{3}$. For solutions with horizon radius greater than
this, no nonconformal negative mode exists. Furthermore, the
correspondence may be extended to any spacetime dimension $n >
4$. This latter observation is of particular relevance to more recent
ideas involving anti-de Sitter spaces.    

Analogous to that found by Gross {\sl et al.}, a single nonconformal
negative mode exists in the $n=5$ dimensions Schwarzschild solution,
which is described by Witten~\cite{WITT1} as leading to semi-classical
instability of the ground state $\mr{M}^4 \times \mr{S}^1$ in the
original Kaluza-Klein theory without fermions. In this case the
periodically identified coordinate is not the imaginary time but the
compact fifth spatial direction. In fact the tunnelling instability
phenomenon, and the associated negative mode, seems to be a quite
general characteristic of any metric with periodic identifications -- 
called a `tachyonic solution'~\cite{GIBB1}. The Lorentzian
continuation of the metric may be interpreted as the result of the
tunnelling instability, or alternatively as providing the Cauchy
development of the time-symmetric initial data set on some plane of
symmetry regarded as $t=0$.

With this in mind, the continuous one-parameter family of initial data
discussed by Horowitz and Myers~\cite{MYERS}, represented by the line
element 
\be
ds^2 = \biggl[ 1 - M/r^2 + r^2/b^2 \biggr] \: d \tau^2 + \biggl[ 1 - M
/ r^2 + r^2/b^2 \biggr] ^{-1} \: dr^2 + r^2 \: d \Omega^2,
\ee
may be considered as the restriction to the equatorial plane of the
three-spheres in the $n=5$ dimensional Schwarzschild-anti-de Sitter 
metric periodically identified in $\tau$. This metric satisfies the
constraint equation $\vphantom{|}^4 R = -12 / b^2$, and the hyper-plane
on which it is defined may be regarded as $t=0$. The radial coordinate
is of course restricted to $r > r_+$ where this is the real positive
root of the metric function as usual. 

The Euclidean section of the full $n=5$ dimensions metric is obtained
by setting $t = iz$, and is now known to possess a single nonconformal
negative mode if $r_+ < b / \sqrt{2}$. This mode will `pass through
zero' at equality, and hence solutions with $r_+ > b / \sqrt{2}$ will
have no negative modes. Furthermore, if $r_+$ is great enough, the
action and hence the energy will be negative. By analogy with the
earlier argument for global thermodynamic stability, this strongly
suggests that, should the action be negative, then indeed it will be a
global minimum for metrics with these boundary conditions.

The idea, then, is that the larger Schwarzschild-anti-de Sitter black
hole solution for a given periodicity $\beta_{\star}$, should its
action be negative, is in fact the global minimum of the action for
periodically identified metrics satisfying the scalar curvature
constraint $R = 2 n \Lambda / (n-2)$. This is clearly very similar to
the proposal in~\cite{MYERS} and it is believed that the above work
provides strong evidence in favour of this conjecture, albeit without
recourse to the AdS/CFT correspondence.

It is a pleasure to acknowledge Stephen Hawking for suggesting this
avenue of research, and to thank Malcolm Perry and Gary Gibbons for
their help and advice during the preparation of this work.


\begin{thebibliography}{99}
 
\bibitem{GPY} D.J. Gross, M.J. Perry, \& L.G. Yaffe, Instability of
Flat Space at Finite Temperature, {\sl Phys Rev} {\bf D25} 330 (1982).

\bibitem{SWH1} S.W. Hawking, Black Hole Thermodynamics, {\sl Phys Rev}
{\bf D13} 191 (1976).

\bibitem{GP1} G.W. Gibbons \& M.J. Perry, Black Holes and Thermal
Green Functions, {\sl Proc R Soc Lond} {\bf A358} 467 (1978).

\bibitem{YORK1} J.W. York, Black Hole Thermodynamics and the
Euclidean Einstein Action, {\sl Phys Rev} {\bf D33} 2092 (1986).

\bibitem{YORK2} B.F. Whiting \& J.W. York, Action Principle and
Partition Function for the Gravitational Field in Black Hole
Topologies, {\sl Phys Rev Lett} {\bf 61} 1336 (1988).

\bibitem{WHITING1} B.F. Whiting, Black Holes and Thermodynamics, in
{\sl Proceedings of the IXth IAMP Congress}, Ed. B. Simon,
I.M. Davies, \& A. Truman (Adam Hilger Ltd., Bristol, 1988). 

\bibitem{ALLEN} B. Allen, Euclidean Schwarzschild Negative Mode, {\sl
Phys Rev} {\bf D30} 1153 (1984). 

\bibitem{BCM} J.D. Brown, J. Creighton, \& R.B. Mann, Temperature,
Energy, and Heat Capacity of Asymptotically Anti-de Sitter Black
Holes, {\sl Phys Rev} {\bf D50} 6394 (1994).

\bibitem{PAGE} S.W. Hawking \& D.N. Page, Thermodynamics of Black
Holes in Anti-de Sitter Space, {\it Commun Math Phys} {\bf 87} 577
(1983). 

\bibitem{MYERS} G.T. Horowitz \& R.C. Myers, The AdS/CFT
Correspondence and a New Positive Energy Conjecture for General
Relativity, {\sl Phys Rev} {\bf D59} 026005 (1999).

\bibitem{WHITING2} B.F. Whiting, Non-Classical Geometries in
Gravitational Thermodynamics, in {\sl Proceedings of the Vth Marcel
Grossman Meeting}, Ed. D.G. Blair \& M.J. Buckingham (Cambridge
University Press, Cambridge, 1988).

\bibitem{YORK3} J.W. York, Quantum Geometry and Gravitational
Thermodynamics, in {\sl Proceedings of the 1989 Chapel Hill Conference
on Grand Unification}, Ed. P.H. Frampton (World Scientific, Singapore,
1989). 

\bibitem{SWH2} S.W Hawking, The Path-Integral Approach to Quantum
Gravity, in {\sl General Relativity: An Einstein Centenary Survey},
Ed. S.W Hawking \& W. Israel (Cambridge University Press, Cambridge,
1979). 

\bibitem{GHP} G.W Gibbons, S.W Hawking, \& M.J. Perry, Path Integrals
and the Indefiniteness of the Gravitational Action, {\sl Nucl Phys}
{\bf B138} 141 (1978).

\bibitem{NUM_REC} W.H. Press, S.A. Teukolsky, W.T. Vetterling, \&
B.P. Flannery, Numerical Recipes in C: The Art of Scientific
Computing, Second Edition, (Cambridge University Press, Cambridge,
1992). 

\bibitem{WITT1} E. Witten, Instability of the Kaluza-Klein Vacuum,
{\sl Nucl Phys} {\bf B195} 481 (1982).

\bibitem{GIBB1} G.W. Gibbons \& D.A. Rasheed, Dyson Pairs and Zero
Mass Black Holes, {\sl Nucl Phys} {B476} 515 (1996).
 
\end{thebibliography}
\end{document}